\documentclass[journal,10pt]{IEEEtran}
\usepackage{graphicx}
\usepackage{subfigure}
\usepackage{float}
\usepackage{subfloat}
\usepackage{amsmath}
\usepackage{amssymb}
\usepackage{amsthm}
\usepackage{epstopdf}
\usepackage{cases}
\usepackage{color}
\usepackage{makecell}
\usepackage{cite}
\usepackage{algorithmic}
\usepackage[ruled]{algorithm2e}

\usepackage[normalem]{ulem}

\makeatletter
\renewcommand\normalsize{%
	\@setfontsize\normalsize\@xpt\@xiipt
	\abovedisplayskip 6.5\p@ \@plus2\p@ \@minus5\p@
	\abovedisplayshortskip \z@ \@plus3\p@
	\belowdisplayshortskip 6.5\p@ \@plus3\p@ \@minus3\p@
	\belowdisplayskip \abovedisplayskip
	\let\@listi\@listI}
\makeatother

\ifCLASSINFOpdf
\else
\fi
\usepackage{caption}
\usepackage{mathtools}

\begin{document}
\setlength{\textfloatsep}{5pt}

%
\title{Sensing-aided Uplink Channel Estimation for Joint Communication and Sensing}
%
%
%

\author{Xu Chen,~\IEEEmembership{Member,~IEEE,}
	Zhiyong Feng,~\IEEEmembership{Senior Member,~IEEE,}
	J. Andrew Zhang,~\IEEEmembership{Senior Member,~IEEE,}\\
	Zhiqing Wei,~\IEEEmembership{Member,~IEEE,}
	Xin Yuan,~\IEEEmembership{Member,~IEEE,}
	and Ping Zhang,~\IEEEmembership{Fellow,~IEEE}
	\thanks{Xu Chen, Z. Feng, and Z. Wei are with Beijing University of Posts and Telecommunications, Key Laboratory of Universal Wireless Communications, Ministry of Education, Beijing 100876, P. R. China (Email:\{chenxu96330, fengzy, weizhiqing\}@bupt.edu.cn).}
	\thanks{J. A. Zhang is with the Global Big Data Technologies Centre, University of Technology Sydney, Sydney, NSW, Australia (Email: Andrew.Zhang@uts.edu.au).}
	\thanks{ Ping Zhang is with Beijing University of Posts and Telecommunications, State Key Laboratory of Networking and Switching Technology, Beijing 100876, P. R. China (Email: pzhang@bupt.edu.cn).}
	\thanks{X. Yuan is with Commonwealth Scientific and Industrial Research Organization (CSIRO), Australia (email: Xin.Yuan@data61.csiro.au).}
	\thanks{Corresponding author: Zhiyong Feng}
}

%
%

\markboth{}%
{Shell \MakeLowercase{\textit{et al.}}: Bare Demo of IEEEtran.cls for IEEE Journals}
%


\maketitle

\newcounter{mytempeqncnt}
\setcounter{mytempeqncnt}{\value{equation}}
\begin{abstract}
The joint communication and sensing (JCAS) technique has drawn great attention due to its high spectrum efficiency by using the same transmit signal for both communication and sensing. Exploiting the correlation between the uplink (UL) channel and the sensing results, we propose a sensing-aided Kalman filter (SAKF)-based channel state information (CSI) estimation method for UL JCAS, which exploits the angle-of-arrival (AoA) estimation to improve the CSI estimation accuracy. A Kalman filter (KF)-based CSI enhancement method is proposed to refine the least-square CSI estimation by exploiting the estimated AoA as the prior information. Simulation results show that the bit error rates (BER) of UL communication using the proposed SAKF-based CSI estimation method approach those using the minimum mean square error (MMSE) method, while at significantly reduced complexity.

\end{abstract}


%
\IEEEpeerreviewmaketitle

\section{Introduction}
%
%
%
%
 Joint communication and sensing (JCAS) is regarded as one of the most promising techniques for dealing with the spectrum congestion problem in the future 6G networks, achieving wireless communication and sensing using the same transmit signals~\cite{liu2020joint, Feng2021JCSC, Saad2020}. The uplink (UL) JCAS estimates sensing parameters from the UL channel state information (CSI)~\cite{Zhang2022ISAC}. Therefore, the CSI is highly related to the environment sensing information such as the angle-of-arrival (AoA), delay (or range), and Doppler frequency for the signals between the base station (BS) and the user. Moreover, the CSI estimation is critical for realizing reliable communications.

The least-square (LS) method has been widely utilized for channel estimation due to its low complexity~\cite{2010MIMO}. Nevertheless, its low CSI estimation accuracy may result in a large signal-to-noise ratio (SNR) loss for communications. To deal with this problem, the minimum mean square error (MMSE) method was proposed to improve the CSI estimation accuracy. However, the high complexity makes it challenging to be utilized in real applications~\cite{Rodger2014principles}. Recently, the deep neural network (DNN)-based estimation method was proposed~\cite{Ge2022}. By using the imperfect CSI to train the CSI estimation DNN, the CSI estimation can be improved. However, this method has high training overhead and thus is not suitable for real-time communication processing.

Exploiting the correlation between the sensing parameters and CSI, we propose a sensing-aided Kalman filter (SAKF)-based CSI estimation method for uplink (UL) JCAS. We propose a Kalman filter (KF)-based CSI enhancement method by exploiting the estimated AoA as the prior information to iteratively suppress the noise terms in CSI estimation. Simulation results show that the proposed SAKF-based CSI estimation method can achieve BER performance approaching that of the MMSE method, while at a significantly reduced complexity. 

\begin{figure}[!t]
	\centering
	\includegraphics[width=0.20\textheight]{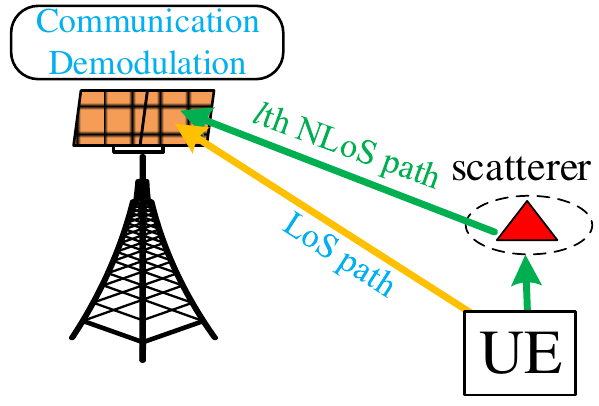}%
	\DeclareGraphicsExtensions.
	\caption{The UL JCAS scenario.}
	\label{fig: Uplink JCS Model}
\end{figure}

\textbf{Notations}: Bold uppercase letters denote matrices (e.g., $\textbf{M}$); bold lowercase letters denote column vectors (e.g., $\textbf{v}$); scalars are denoted by normal font (e.g., $\gamma$); the entries of vectors or matrices are referred to with square brackets; $\left(\cdot\right)^{*}$ and $\left(\cdot\right)^T$ denote Hermitian transpose, complex conjugate and transpose, respectively; ${\bf M}_1 \in \mathbb{C}^{M\times N}$ and ${\bf M}_2 \in \mathbb{R}^{M\times N}$ are ${M\times N}$ complex-value and real-value matrices, respectively; and $v \sim \mathcal{CN}(m,\sigma^2)$ means $v$ follows a complex Gaussian distribution with mean $m$ and variance $\sigma^2$.

\section{System Model}\label{sec:system-model}

We consider a UL JCAS system, where the BS and the user are equipped with uniform plane arrays (UPAs), as shown in Fig.~\ref{fig: Uplink JCS Model}. In the UL preamble (ULP) period, the user transmits the UL preamble signals, and BS receives them for CSI estimation, which is further used for estimating the sensing parameters, such as the AoAs, ranges and Doppler shifts. In the UL data (ULD) period, the BS receives and demodulates the UL data signals from the user using the estimated CSI. 

Next, we introduce the transmit signal and received signal models.

\subsection{Transmit Signal Model} \label{subsec:DUC_signal model}
Orthogonal frequency division multiplexing (OFDM) signal is adopted as the transmit signal, which is given by
\begin{equation}\label{equ:DUC_signal}
	{s}( t ) = \sum\limits_{m = 0}^{M_s - 1} {\sum\limits_{n = 0}^{N_c - 1} {\sqrt {P_t^U} d_{n,m}{e^{j2\pi ( {{f_c} + n\Delta {f}} )t}}} } {\mathop{\rm Rect}\nolimits} \left(\frac{{t - mT_s}}{{T_s}}\right),
\end{equation}
where ${P_t^U}$ is the transmit power, ${M_s}$ and ${N_c}$ are the numbers of OFDM symbols and subcarriers, respectively; ${d_{n,m}}$ is the transmit OFDM baseband symbol of the $m$th OFDM symbol at the $n$th subcarrier, $f_c$ is the carrier frequency, $\Delta f$ is the subcarrier interval, ${T_s} = \frac{1}{{\Delta f}} + {T_g}$ is the time duration of each OFDM symbol, and ${T_g}$ is the guard interval.

\begin{figure}[!t]
	\centering
    \includegraphics[width=0.22\textheight]{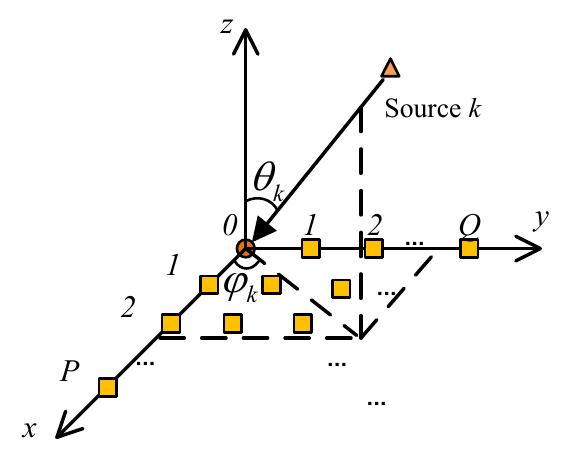}%
	\DeclareGraphicsExtensions.
	\caption{The UPA model.}
	\label{fig: UPA model}
\end{figure}

\subsection{UPA Model} \label{subsec:UPA model}
Fig.~\ref{fig: UPA model} demonstrates the UPA model. The uniform interval between neighboring antenna elements is denoted by $d_a$. The size of UPA is ${P} \times {Q}$. The AoA for receiving or the angle-of-departure (AoD) for transmitting the $k$th far-field signal is ${{\bf{p}}_k} = {( {{\varphi _k},{\theta _k}} )^T}$, where ${\varphi _k}$ and ${\theta _k}$ are the azimuth and elevation angles, respectively. The phase difference between the ($p$,$q$)th antenna element and the reference antenna element is 
\begin{equation}\label{equ:phase_difference}
	{a_{p,q}}\left( {{{\bf{p}}_k}} \right) \! =\! \exp [ { - j\frac{{2\pi }}{\lambda }{d_a}( {p\cos {\varphi _k}\sin {\theta _k} \!+\! q\sin {\varphi _k}\sin {\theta _k}} )} ],
\end{equation}
where $\lambda = c/f_c$ is the wavelength of the carrier, $f_c$ is the carrier frequency, and $c$ is the speed of light. The steering vector for the array is given by 
\begin{equation}\label{equ:steeringVec}
	{\bf{a}}\left( {{{\bf{p}}_k}} \right) = [ {{a_{p,q}}\left( {{{\bf{p}}_k}} \right)} ]\left| {_{p = 0,1,\cdots,P - 1;q = 0,1,\cdots,Q - 1}}\right.,
\end{equation}
where $\mathbf{a}(\mathbf{p}_k) \in \mathbb{C}^{PQ \times 1}$, and ${ {[ {{v_{p,q}}} ]} |_{(p,q) \in {\bf{S}}1 \times {\bf{S}}2}}$ denotes the vector stacked by values ${v_{p,q}}$ satisfying $p\in{\bf{S}}1$ and $q\in{\bf{S}}2$. The sizes of the antenna arrays of the BS and the user are ${P_t} \times {Q_t}$ and ${P_r} \times {Q_r}$, respectively.

\subsection{Received Signal Model}
The frequency-domain received communication signal at the $n$th subcarrier of the $m$th OFDM symbol is expressed as~\cite{Zhang2022ISAC} 
\begin{equation}\label{equ:y_C^U}
\begin{array}{l}
	{{\bf{y}}_{C,n,m}}\!\! =\!\! \sqrt {P_t^U} {d_{n,m}}\!\!\sum\limits_{l = 0}^{L - 1} {\!\! \left[\!\! \begin{array}{l}
			{e^{j2\pi m{T_s}[{f_{c,d,l}} + {\delta _f}\left( m \right)]}} \\
			{e^{ - j2\pi n\Delta f[{\tau _{c,l}} + {\delta _\tau }\left( m \right)]}}\\
			\times {b_{C,l}} {\chi _{TX,l}}{\bf{a}}( {{\bf{p}}_{RX,l}^U})
		\end{array} \!\! \right]} \!+\! {{\bf{n}}_{t,n,m}},
\end{array}
\end{equation}
where $l = 0$ is for the channel response of the LoS path, and $l \in \{1, \cdots, L-1\}$ is for the $l$th non-LoS (NLoS) path; $\chi _{TX,l} = {{\bf{a}}^T}( {{\bf{p}}_{TX,l}^U} ) {{\bf{w}}_{TX}}$ is the transmit BF gain, ${\bf{w}}_{TX}\in \mathbb{C}^{P_r Q_r \times 1}$ is the transmit beamforming~(BF) vector, and ${\bf{a}}( {{\bf{p}}_{TX,l}^U} ) \in \mathbb{C}^{{P_r}{Q_r}} \times 1$ and ${\bf{a}}( {{\bf{p}}_{RX,l}^U} ) \in \mathbb{C}^{{P_t}{Q_t}} \times 1$ are the steering vectors for UL transmission and receiving, respectively; ${f_{c,d,l}}$ and $\tau_{c,l}$ are the Doppler and range of the $l$th path, respectively; ${\delta _f}\left( m \right)$ and ${\delta _\tau }\left( m \right)$ are the frequency and timing offsets, respectively; $b_{C,0}$ = 1, $b_{C,l}$ ($l>0$) is a random variable following $\mathcal{CN}(0,\sigma _{\beta_l}^2)$; ${\bf{n}}_{t,n,m}$ is the combined noise including Gaussian noise and possible reflected interferences, and each element of
${\bf{n}}_{t,n,m}$ follows $\mathcal{CN}(0,\sigma _N^2)$; and ${\bf{y}}_{C,n,m}, {\bf{n}}_{t,n,m} \in \mathbb{C}^{{P_t}{Q_t} \times 1}$. We adopt the low-complexity LS method to generate ${\bf{w}}_{TX}$, i.e., ${\bf{w}}_{TX} = {[ {{{\bf{a}}^T}( {\tilde{\bf{p}}_{TX,l}^U} )} ]^\dag }$~\cite{Zhang2019JCRS}, where $[\textbf{A}]^\dag$ is the pseudo-inverse matrix of $\textbf{A}$. When the transmit beam alignment is complete, ${\tilde{\bf{p}}_{TX,l}^U} \approx {{\bf{p}}_{TX,l}^U}$.
We denote the actual CSI as
\begin{equation}\label{equ:h_c_U}
	{\bf{h}}_{C,n,m} = \sum\limits_{l = 0}^{L - 1} \!\! \left[\!\! \begin{array}{l}
			{e^{j2\pi m{T_s}[{f_{c,d,l}} + {\delta _f}\left( m \right)]}} {e^{ - j2\pi n\Delta f[{\tau _{c,l}} + {\delta _\tau }\left( m \right)]}}\\
			\times {b_{C,l}} \sqrt{P_t^U} {\chi _{TX,l}}{\bf{a}}( {{\bf{p}}_{RX,l}^U})
		\end{array} \!\! \right].
\end{equation}

\begin{figure}[!t]
	\centering
	\includegraphics[width=0.37\textheight]{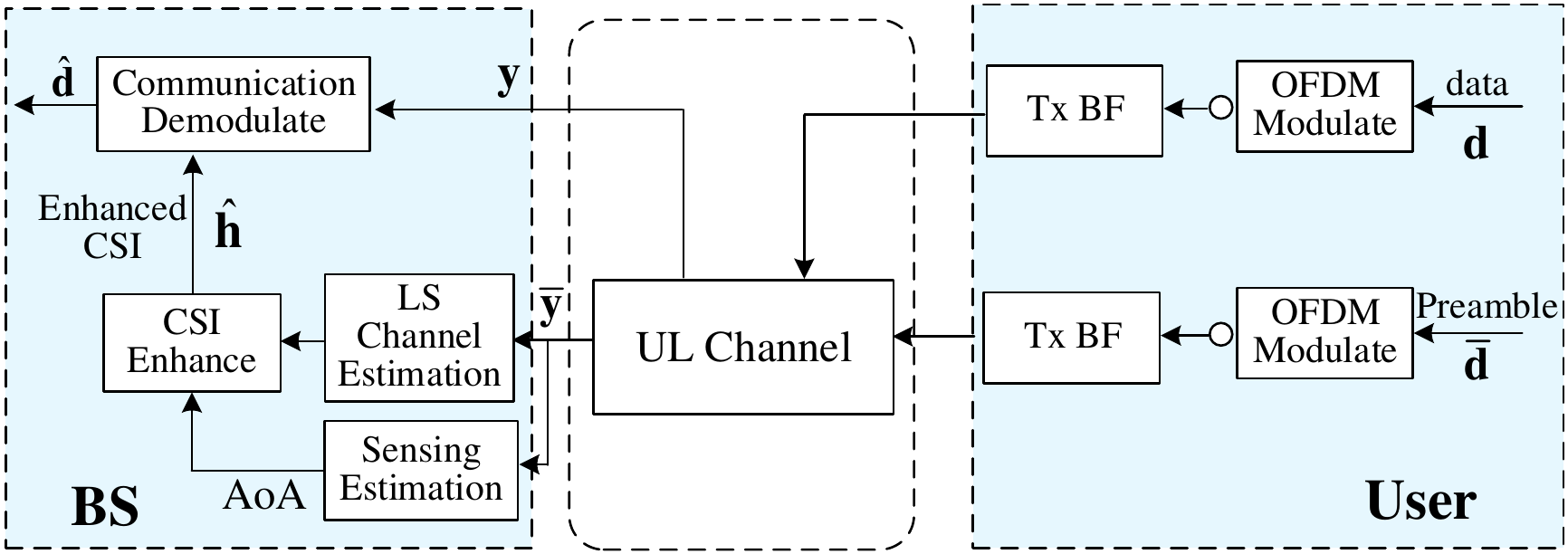}%
	\DeclareGraphicsExtensions.
	\caption{The UL JCAS scheme with the SAKF-based CSI estimation.}
	\label{fig: UL communication}
\end{figure}

\section{SAKF-based CSI Estimation Method}\label{sec:JCAS_signal_processing}
In this section, we present the SAKF-based CSI estimation method. The UL JCAS scheme is shown in Fig.~\ref{fig: UL communication}.
In the ULP period, the UL CSI estimation at the $n$th subcarrier of the $m$th OFDM symbol is obtained with the LS method as~\cite{2010MIMO}
\begin{equation}\label{equ:h_c_U_bar}
		{{{\bf{\hat h}}}_{C,n,m}} = \frac{{{{\bf{y}}_{C,n,m}}}}{{{d_{n,m}}}} = {{\bf{h}}_{C,n,m}} + {{{\bf{\bar n}}}_{t,n,m}} \in {\mathbb{C}^{{P_t}{Q_t} \times 1}}
\end{equation}
where ${{{\bf{y}}_{C,n,m}}}$ is given in \eqref{equ:y_C^U}, and ${d_{n,m}}$ is the preamble symbol with unit constant modulus. Since $N_c$ subcarriers at $M_s$ OFDM preamble symbols are used, we can stack all the CSIs to obtain the matrix ${\bf{\hat H}}_C \in {\mathbb{C}^{{P_t}{Q_t} \times N_c M_s}}$, where the $[(m-1)N_c + n]$th column of ${\bf{\hat H}}_C$ is ${{\bf{\hat h}}_{C,n,m}}$. 

In UL JCAS operation, the AoAs, ranges, and Dopplers are jointly estimated, as discussed in \cite{2021YangJEVAR}. Specifically, the number of incident signals and their AoAs can be jointly estimated with the minimum description length (MDL)-based multiple signal classification (MUSIC) method~\cite{2017GAOMDLMUSIC}. Besides, the noise power can be estimated by averaging the last $P_t Q_t-L$ eigenvalues of ${{\bf{\hat H}}_C}{( {{{{\bf{\hat H}}}_C}} )^H}$ as $\hat \sigma_N^2$.

The AoA can be estimated once and used over multiple consecutive packets. According to \eqref{equ:h_c_U_bar}, we can see that ${{\bf{\hat h}}_{C,n,m}}{( {{{{\bf{\hat h}}}_{C,n,m}}} )^H}$ does not contain ${e^{j2\pi m{T_s}[{f_{c,d,l}} + {\delta _f}\left( m \right)]}}{e^{ - j2\pi n\Delta f[{\tau _{c,l}} + {\delta _\tau }\left( m \right)]}}$. Therefore, the AoA estimation is not prominently affected by the frequency and timing offset, while it is a challenging issue for estimating the range and Doppler~\cite{Zhang2022ISAC}. Thus, AoA is the most suitable sensing parameter to refine the communication CSI. Next, we propose the KF-based CSI enhancement method that uses the estimated AoAs to refine the CSI estimation.

\subsection{KF-based CSI Enhancement Method}\label{sec:UplinkJCAS}
Reshape ${{\bf{\hat h}}_{C,n,m}}$ and ${{\bf{h}}_{C,n,m}}$ into ${\bf{\hat H}}_{C,n,m}$ and ${{\bf{H}}_{C,n,m}} \in \mathbb{C}^{{P_t} \times {Q_t}}$, respectively. According to \eqref{equ:y_C^U}, if we treat the $l$th incident signal as the observing signal, then ${[ {{\bf{\hat H}}_{C,n,m}} ]_{p,q}}$ can be rewritten as
\begin{equation}\label{equ:H_c_U}
	\begin{aligned}
		&{[ {{\bf{\hat H}}_{C,n,m}} ]_{p,q}} = {[ {{\bf{H}}_{C,n,m}^{l}} ]_{p,q}} + {[ {{\bf{I}}_{n,m}^{l}} ]_{p,q}} + n_{t,n,m}^{p,q}\\
		&= \sqrt {P_t^U} \alpha _{n,m,l}{e^{ - j\frac{{2\pi }}{\lambda }{d_a}( {p\cos {\varphi _l}\sin {\theta _l} + q\sin {\varphi _l}\sin {\theta _l}} )}} \\ & \quad + {[ {{\bf{I}}_{n,m}^{l}} ]_{p,q}} + n_{t,n,m}^{p,q},
	\end{aligned}
\end{equation}
where ${\alpha _{n,m,l}} = {b_{C,l}}{e^{j2\pi mT_s{{\tilde f}_{d,l,m}}}}{e^{ - j2\pi n\Delta f{{\tilde \tau }_{l,m}}}}{\chi _{TX,l}}$, ${\tilde f_{d,l,m}} = {f_{c,d,l}} + {\delta _f}\left( m \right)$, ${\tilde \tau _{l,m}} = {\tau _{c,l}} + {\delta _\tau }\left( m \right)$, and $n_{t,n,m}^{p,q} = {\left[ {{{{\bf{\bar n}}}_{t,n,m}}} \right]_{p{Q_t} + q}}$. Moreover, $[{{\bf{I}}_{n,m}^l}]_{p,q}$ is the interference signals in other paths, and is expressed as
\begin{equation}\label{equ:I_nm}
	{\left[ {{\bf{I}}_{n,m}^l} \right]_{p,q}}\!\!\! = \!\! \sqrt {P_t^U}\!\!\!\! \sum\limits_{i = 0,i \ne l}^{L - 1}\!\!\!\!\! { {{\alpha _{n,m,i}}{e^{ - j\frac{{2\pi{d_a} }}{\lambda }\left( {p\cos {\varphi _i}\sin {\theta _i} \!+ q\sin {\varphi _i}\sin {\theta _i}} \right)}}} }.
\end{equation}

According to \eqref{equ:H_c_U}, we can treat ${[ {{\bf{\hat H}}_{C,n,m}} ]_{p,q}}$ as the noisy observation of ${[ {{\bf{H}}_{C,n,m}^{l}} ]_{p,q}}$.  The state transfer expressions of ${\bf{H}}_{C,n,m}^l$ in the $p$-axis and $q$-axis are expressed, respectively, as
\begin{equation}\label{equ:state_transfer_p}
	{\left[ {{\bf{H}}_{C,n,m}^l} \right]_{p + 1,q}} = {\left[ {{\bf{H}}_{C,n,m}^l} \right]_{p,q}}{A_{P,l}},
\end{equation}
\begin{equation}\label{equ:state_transfer_q}
	{\left[ {{\bf{H}}_{C,n,m}^l} \right]_{p,q + 1}} = {\left[ {{\bf{H}}_{C,n,m}^l} \right]_{p,q}}{A_{Q,l}},
\end{equation}
where $A_{P,l}$ and $A_{Q,l}$ are the transfer factors. Using the estimated ${\bf{\hat p}}_{RX,l}^U = ( {{{\hat \varphi }_l},{{\hat \theta }_l}} )$, we obtain the estimations of $A_{P,l}$ and $A_{Q,l}$ as
\begin{equation}\label{equ:Ap_Aq}
	{\hat A_{P,l}} = {e^{ - j\frac{{2\pi }d_a}{\lambda }\cos {\hat \varphi _l}\sin {\hat \theta _l}}},
	{\hat A_{Q,l}} = {e^{ - j\frac{{2\pi d_a}}{\lambda }\sin {\hat \varphi _l}\sin {\hat \theta _l}}}.
\end{equation}
According to \eqref{equ:state_transfer_p} and \eqref{equ:state_transfer_q}, each row and column of ${{\bf{\hat H}}_{C,n,m}}$ can be filtered by a KF filter to suppress the noise in \eqref{equ:H_c_U}. Let ${{\bf{\hat h}}_C}$ and ${{\bf{\bar h}}_C}$ denote a row or a column of ${{\bf{\hat H}}_{C,n,m}}$ and its filtered vector, respectively. The prior estimation of ${[ {{{{\bf{\hat h}}}_C}} ]_p}$ can be expressed as
\begin{equation}\label{equ:hc_p}
	[ {{{{\bf{\hat h}}}_C}} ]_p^ -  = {[ {{{{\bf{\hat h}}}_C}} ]_{p - 1}}A, for \; p \in \{ 1, \cdots P - 1\},
\end{equation}
where $A = $ ${\hat A_{P,l}}$ or ${\hat A_{Q,l}}$ when ${[ {{{{\bf{\hat h}}}_C}} ]_p}$ is a column or a row vector, respectively. Then, ${[ {{{{\bf{\bar h}}}_C}} ]_p}$ can be further updated as~\cite{2017Kalman}
\begin{equation}\label{equ:hc_p_bar}
	{[ {{{{\bf{\bar h}}}_C}} ]_p}{\rm{ = }}[ {{{{\bf{\hat h}}}_C}} ]_p^ - {\rm{ + }}{K_p}( {{{[ {{{{\bf{\hat h}}}_C}} ]}_p} - [ {{{{\bf{\hat h}}}_C}} ]_p^ - } ),
\end{equation}
where $K_p$ is the data fusion factor. Moreover, ${K_p}$ is expressed as~\cite{2017Kalman}
\begin{equation}\label{equ:K_p}
    \begin{array}{ll}
         {K_p} = {\left( {p_{w,p}^ - } \right)^*}{(p_{w,p}^ -  + \sigma _N^2)^{ - 1}},\\
        p_{w,p}^ -  = A{p_{w,p - 1}}{A^*},\\
        {p_{w,p}} = \left( {1 - {K_p}} \right)p_{w,p}^{-},
    \end{array}
\end{equation}
where $\sigma _N^2$ is the power of the interference-plus-noise, its estimation is $\hat \sigma_N^2$, and ${p_{w,0}}$ is the variance of initial observation. Based on \eqref{equ:state_transfer_p} and \eqref{equ:state_transfer_q}, we obtain 
\begin{equation}\label{equ:p_w0}
    {p_{w,0}} = \frac{1}{P}\sum\limits_{p = 0}^{P - 1} {|{{[{{{\bf{\hat h}}}_C}]}_p}{{\left( A \right)}^{ - p}} - {{[{{{\bf{\hat h}}}_C}]}_0}|_2^2}.
\end{equation}
Based on \eqref{equ:hc_p}, \eqref{equ:hc_p_bar}, \eqref{equ:K_p}, and \eqref{equ:p_w0}, we propose the KF-based CSI enhancement method as shown in \textbf{Algorithm~\ref{Kalman_CSI}}. Note that we further add an inverse version of KF in \textbf{Step} 5 to completely exploit the sensing information, where the transfer factor is updated as ${A^{ - 1}}$. By exploiting the estimated AoAs of JCAS as the prior information, \textbf{Algorithm~\ref{Kalman_CSI}} can suppress the noise terms in \eqref{equ:H_c_U}.

\begin{algorithm}[!t]
	\caption{KF-based CSI Enhancement method}
	\label{Kalman_CSI}
	\KwIn{The observation variance $\sigma _N^2 = \hat \sigma _N^2$; The transfer factor $A = {\hat A_p}$; The observation sequence ${{\bf{\hat h}}_C} = {[ {{\bf{\hat H}}_{C,n,m}^l} ]_{:,q}}$.
	}
	\KwOut{Filtered sequence ${{\bf{\bar h}}_C} = {[ {{\bf{\bar H}}_{C,n,m}^l} ]_{:,q}}$.}
	\textbf{Step} 1: The dimension of ${{\bf{\hat h}}_C}$ is obtained as $P$;
	
	\textbf{Step} 2: ${[ {{{{\bf{\bar h}}}_C}} ]_0} = {[{\bf{\hat h}}_C]_{0}}$; 
	
	\textbf{Step} 3: ${p_{w,0}} = \frac{1}{P}
	{{\sum\limits_{p = 0}^{P - 1} {| {{{[ {{{{\bf{\hat h}}}_C}} ]}_p}{{\left( A \right)}^{ - p}} - {{[ {{{{\bf{\hat h}}}_C}} ]}_0}} |_2^2} } }$; 
	
	\textbf{Step} 4: \\
	\For{$p$ = {\rm 1} to $P - 1$} {
		$[ {{{{\bf{\hat h}}}_C}} ]_p^ -  = A{[ {{{{\bf{\bar h}}}_C}} ]_{p - 1}}$\;
		$p_{w,p}^ -  = A{p_{w,p - 1}}{A^*}$\;
		${K_p} = {\left( {p_{w,p}^ - } \right)^*}{( {p_{w,p}^ -  + \sigma _N^2} )^{ - 1}}$\;
		${[ {{{{\bf{\bar h}}}_C}} ]_p} = [ {{{{\bf{\hat h}}}_C}} ]_p^ -  + {K_p}( {{{[ {{{{\bf{\hat h}}}_C}} ]}_p} - [ {{{{\bf{\hat h}}}_C}} ]_p^ - } )$\;
		${p_{w,p}} = \left( {1 - {K_p}} \right)p_{w,p}^ - $\;
	}
	\textbf{Step} 5:
	\For{$p$ = $P-1$ to 1} {
		$[ {{{{\bf{\hat h}}}_C}} ]_{p - 1}^ -  = {A^{ - 1}}{[ {{{{\bf{\bar h}}}_C}} ]_p}$\;
		$p_{w,p - 1}^ -  = {A^{ - 1}}{p_{w,p}}{( {{A^{ - 1}}} )^*}$\;
		${K_p} = {( {p_{w,p - 1}^ - } )^*}{( {p_{w,p - 1}^ -  + \sigma _N^2} )^{ - 1}}$\;
		${[ {{{{\bf{\bar h}}}_C}} ]_{p - 1}} = [ {{{{\bf{\hat h}}}_C}} ]_{p - 1}^ -  + {K_p}( {{{[ {{{{\bf{\hat h}}}_C}} ]}_{p - 1}} - [ {{{{\bf{\hat h}}}_C}} ]_{p - 1}^ - } )$\;
		${p_{w,p - 1}} = \left( {1 - {K_p}} \right)p_{w,p - 1}^ - $\;
	}
	\Return ${ {[ {\bar h_{C,n,m}} ]} |_{n = 0, \cdots ,N_c - 1}}$.
\end{algorithm}

By using \textbf{Algorithm~\ref{Kalman_CSI}} to filter $Q_t$ columns of ${\bf{\hat H}}_{C,n,m}$ in parallel with ${\hat A_{P,l}}$, we can obtain ${\bf{\hat H}}_{C,n,m}^{l,(1)}$. Further, we can also use \textbf{Algorithm~\ref{Kalman_CSI}} to filter the $p$th row of ${\bf{\hat H}}_{C,n,m}^{l,(1)}$ by replacing the input with $A = {\hat A_{Q,l}}$ and ${{\bf{\hat h}}_C} = {[ {{\bf{\hat H}}_{C,n,m}^{l,(1)}} ]_{p,:}}$. After filtering all $P_t$ rows of ${\bf{\hat H}}_{C,n,m}^{l,(1)}$, we obtain ${\bf{\hat H}}_{C,n,m}^{l,(2)}$. After obtaining all the ${\bf{\hat H}}_{C,n,m}^{l,(2)}$ for $l = 0, \cdots, L-1$, the aggregation of all $L$ channel components is given by ${\bf{\hat H}}_{C,n,m}^{(2)} = \sum\limits_{l = 0}^{L - 1} {{\bf{\hat H}}_{C,n,m}^{l,(2)}}$.

After being filtered by \textbf{Algorithm~\ref{Kalman_CSI}}, the noise term in the CSI estimation will be suppressed. Vectorize ${\bf{\hat H}}_{C,n,m}^{(2)}$ as ${\bf{\hat h}}_{C,n,m}^{(2)} = vec( {{\bf{\hat H}}_{C,n,m}^{(2)}} ) \in {\mathbb{C}^{{P_t}{Q_t} \times 1}}$. Then, we can use ${\bf{\hat h}}_{C,n,m}^{(2)}$ to demodulate UL data signals with zero-forcing (ZF) receive BF~\cite{2010MIMO}.

\subsection{Complexity Analysis}
In this subsection, we analyze the complexity of the above SAKF-based CSI estimation and compare it with the LS and MMSE CSI estimation methods. The CSI estimation of the MMSE method can be expressed as~\cite{2010MIMO}
\begin{equation}\label{equ:MMSE}
	{{\bf{\hat h}}_{MMSE}} = {{\bf{R}}_{{\bf{hh}}}}{\left[ {{{\bf{R}}_{{\bf{hh}}}} + \hat \sigma _N^2{\bf{I}}} \right]^{ - 1}}{{\bf{\hat h}}_{C,n,m}},
\end{equation}
where ${{\bf{R}}_{{\bf{hh}}}} = E\left( {{{\bf{h}}_{C,n,m}}{\bf{h}}_{C,n,m}^H} \right)$.

Let $N = {P_t}{Q_t}$. The complexity of the LS method comes from the complex-value division, which is $\mathcal{O}(N)$. The MMSE method adds the matrix inverse and multiplication operations based on the LS method. Therefore, the complexity of the MMSE method is $\mathcal{O}(N^3)$. In contrast, the SAKF method adds scalar KF iterations with only two rounds of circulations and uses the estimated AoAs of JCAS without additional sensing processing. Thus, the complexity of the SAKF method for JCAS system is $\mathcal{O}(N+2N) = \mathcal{O}(3N)$.

\section{Simulation Results}\label{sec:Simulation}
In this section, we present the simulation results of the BERs using the proposed SAKF method, compared with the LS and MMSE methods. The simulation parameters are listed as follows.

The carrier frequency is set to 28 GHz~\cite{3GPPV2X}, the antenna interval, $d_a$, is half the wavelength, the sizes of antenna arrays of the BS and user are $P_t \times Q_t = 8 \times 8$ and $P_r \times Q_r = 1\times 1$, respectively, and the number of paths is $L = 2$. The subcarrier interval is $\Delta {f} =$ 480 kHz, the subcarrier number is $N_c =$ 256, and the bandwidth is ${{B  =  }}{N_c}\Delta f = $122.88 MHz. The number of OFDM symbols used for simulation is $M_s = 64$.
The variance of the Gaussian noise is $\sigma_N^2 = 4.9177\times10^{-12} $ W.
The transmit power is determined according to the given SNR and  $\sigma_N^2$. UL SNR is defined as the SNR of each antenna element of BS. According to \eqref{equ:y_C^U}, the UL SNR is expressed as${\gamma _c} = {{P_t^U\sum\limits_{l = 0}^{L - 1} {{{\left| {{b_{C,l}}\chi _{TX,l}^U} \right|}^2}} } \mathord{\left/
		{\vphantom {{P_t^U\sum\limits_{l = 0}^{L - 1} {{{\left| {{b_{C,l}}\chi _{TX,l}^U} \right|}^2}} } {\sigma _N^2}}} \right.
		\kern-\nulldelimiterspace} {\sigma _N^2}}$.

\begin{figure}[!t]
	\centering
	\includegraphics[width=0.30\textheight]{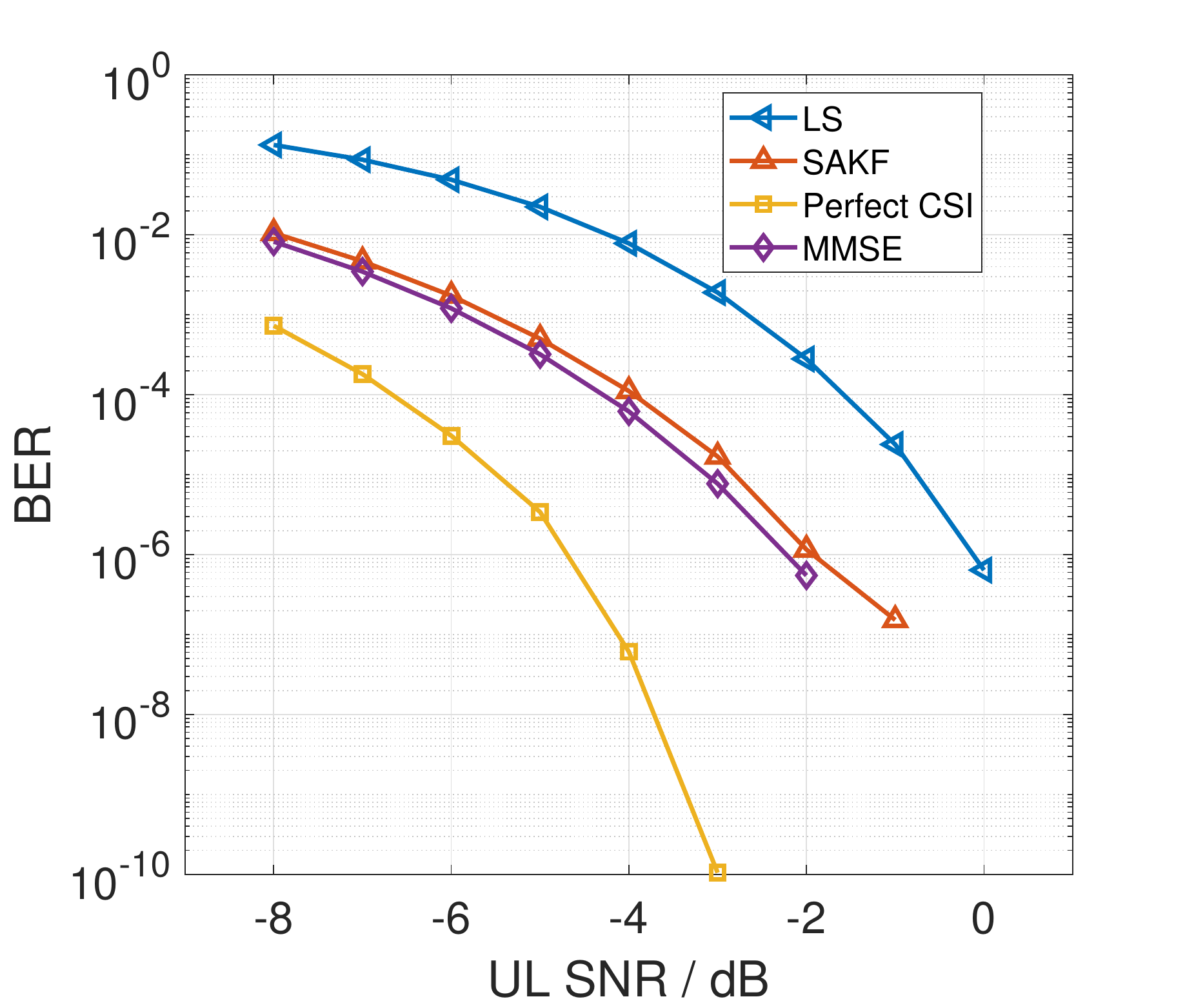}%
	\DeclareGraphicsExtensions.
	\caption{The BERs using the SAKF, LS, and MMSE methods under 4-QAM modulation.}
	\label{fig: BER_4QAM_whole}
\end{figure}

Fig.~\ref{fig: BER_4QAM_whole} demonstrates the BERs using the SAKF, LS, and MMSE methods under 4-QAM modulation. The required SNR to achieve the given BER for the SAKF method is about~1.8 dB lower than that for the LS method, but 0.2 dB higher than that for the MMSE method. This is because \textbf{Algorithm~\ref{Kalman_CSI}} filters the CSI estimated by the LS method exploiting the estimated AoAs.

\begin{figure}[!t]
	\centering
	\includegraphics[width=0.30\textheight]{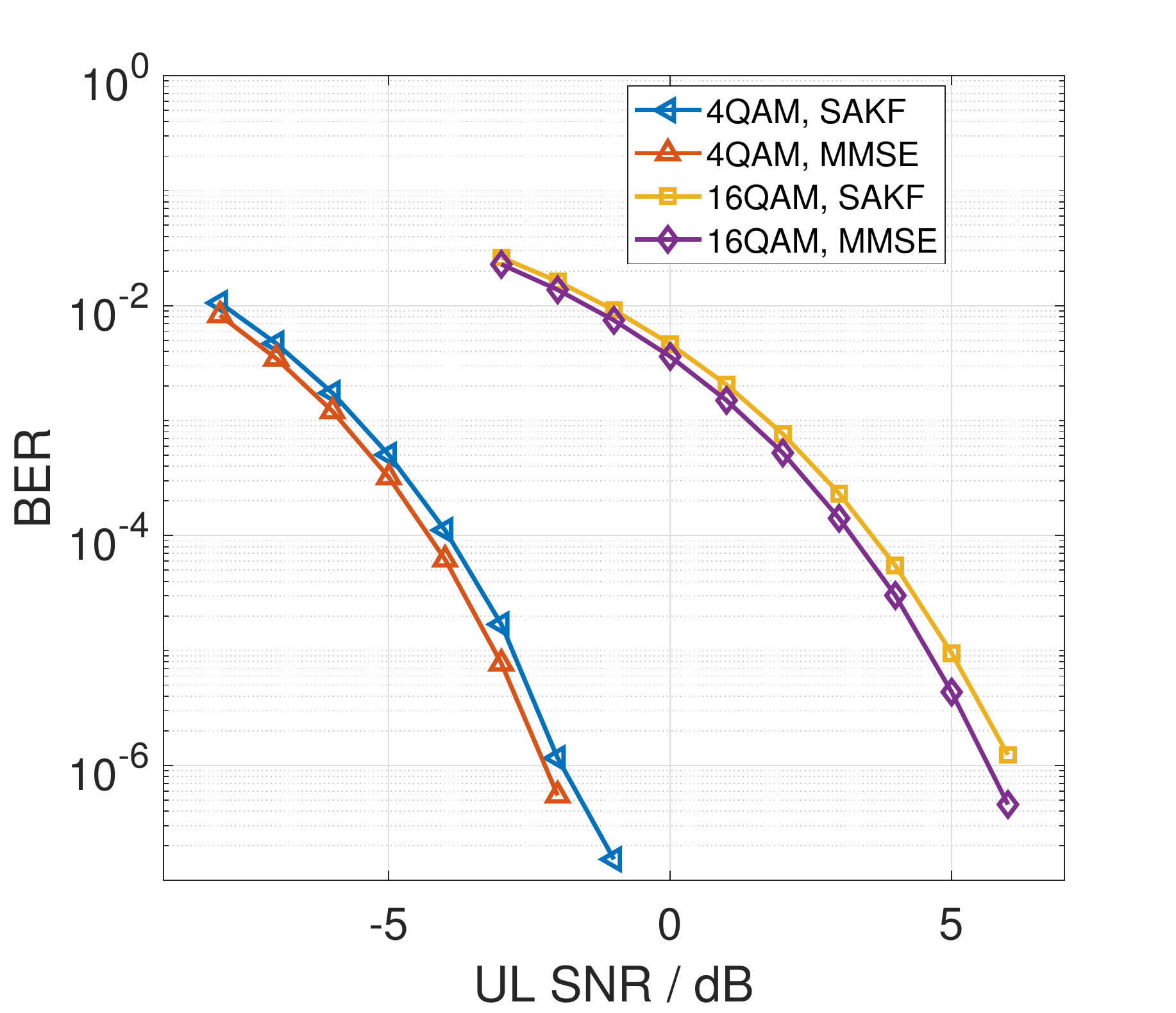}%
	\DeclareGraphicsExtensions.
	\caption{The BERs using the SAKF and MMSE methods under 4-QAM and 16-QAM modulations.}
	\label{fig: BER_4_16QAM}
\end{figure}

Fig.~\ref{fig: BER_4_16QAM} plots the BER curves using the SAKF and MMSE methods under 4-QAM and 16-QAM modulations. It can be seen that the BER performance of the proposed SAKF method can approach that of the MMSE method in both low and high QAM orders. The above results indicate that the proposed SAKF method can approach the MMSE method in BER performance with the significantly reduced complexity.

\section{Conclusion}\label{sec:conclusion}
In this paper, we propose a SAKF-based UL channel estimation method for JCAS system. The KF exploits the AoAs estimated from preamble signals as the prior information to refine the LS CSI estimation. Simulation results show that the SNR required to achieve the given BER for the proposed SAKF method is about 1.8 dB lower than the LS method, and about 0.2 dB higher than the MMSE method. 



%

{\small
	\bibliographystyle{IEEEtran}
	\bibliography{reference}
}
\vspace{-10 mm}
\ifCLASSOPTIONcaptionsoff
  \newpage
\fi

\end{document}